\documentclass[]{aastex6}

\usepackage{multirow}
\fullcollaborationName{The Friends of AASTeX Collaboration}

\begin{document}
 \fontsize{10pt}{10pt}{Comments: Accepted for publication in The Astrophysical Journal. 7 pages, 8 figures, 3 tables.}
\title{likely Detection of GeV gamma-Ray Emission from the composite supernova remnant COMP G327.1+1.1 with Fermi-LAT}

\author{Yunchuan Xiang\altaffilmark{1}, Yi Xing\altaffilmark{2} and Zejun Jiang\altaffilmark{1}}

\altaffiltext{1}{Department of Astronomy, Yunnan University, and Key Laboratory of Astroparticle Physics of Yunnan Province, Kunming, 650091, China, xiang{\_}yunchuan@yeah.net, zjjiang@ynu.edu.cn} 

\altaffiltext{2}{Key Laboratory for Research in Galaxies and Cosmology,
Shanghai Astronomical Observatory, Chinese Academy of Sciences,
80 Nandan Road, Shanghai, 200030, China}


\begin{abstract}
We report the likely GeV $\gamma$-ray emission from the composite supernova remnant (SNR) COMP G327.1+1.1 by analyzing $\sim$12.2 years of the \textit{Fermi} Large Area Telescope (\textit{Fermi}-LAT) Pass 8 data.
We found the features of its \textbf{ranges of} spectrum and luminosity are well consistent with those of the observed COMP SNRs in the Milky Way.
Moreover, the position of the source matches those in radio and TeV energy bands, 
we propose that the $\gamma$-ray source is likely to be a  GeV counterpart of COMP G327.1+1.1. 
Considering the case of the association from COMP G327.1+1.1 and the $\gamma$-ray source,
we interpreted its broadband spectral energy distribution (SED) by using three simple stationary models including one-zone and two-zone leptonic models and one-zone leptohadronic model.
We found that the simple two-zone model dominated by leptons can better explain its SED.
 More high-energy data are expected to firmly confirm the association between the $\gamma$-ray source and COMP G327.1+1.1 in the future.
\end{abstract}

\section{Introduction} \label{sec:intro}
COMP G327.1+1.1, as a non-thermal radio source, was reported by 
\citet{Clark1973,Clark1975}.
Its mean angular diameter was 16.3$^{'}$, and its flux density was 10.6 Jy at 408 MHz, and that at 5000 MHz was 4.3 Jy \citep{Clark1973}. The  emission at 408 MHz was described as a shell source with a strong peak, and that of 5000 MHz also showed a similar peak, but did not exhibit a shell structure.
At 843 MHz band observed by the Molonglo Observatory Synthesis Telescope (MOST), its flux was 7.6 Jy and  was identified as a composite SNR with an unusual off-centre plerionic component and a faint shell, with the flux density of the unusual component being 2 Jy \citep{Whiteoak1996}.

Initially, X-ray detection was reported by the Einstein Observatory
\citep{Lamb1981}. Its energy flux of 1-3 keV was (0.81$\pm$0.12) $\times 10^{-12}$ erg $\rm cm^{-2}$ $\rm s^{-1}$, and its luminosity was (0.73$\pm$0.11) $\times 10^{-34}$ erg $\rm s^{-1}$.
For 0.7-2.2 keV from the \textit{ROSAT} PSPC observations,  
40\% of the X-ray counts came from the brighter central region, and the spectrum of the central region was harder than that of the shell of the south \citep{Seward1996, Sun1999}.
Based on the observation data from ROSAT and Advanced Satellite for Cosmology and Astrophysics (ASCA), \citet{Sun1999} considered that COMP G327.1+1.1 contained a non-thermal emission component and a diffuse thermal emission component. The non-thermal emission component was from the synchrotron nebula powered by an undiscovered central pulsar, and the thermal component was from the interaction between the shock and interstellar medium (ISM). 
Using the X-ray observation from BeppoSAX, \citet{Bocchino2003} studied the origin of its X-ray emission, 
 they considered COMP G327.1-1.1 went through a long evolutionary time than previously estimated (1.1 $\times$ 10$^{4}$ years) according to the theoretical model of Sedov or radiative expansion.
The X-ray observatories of Chandra and XMM-Newton show a pulsar wind nebula (PWN) inside COMP G327.1-1.1 according to an unusual morphology \citep{Temim2009}. 

In the TeV energy band, \citet{Acero2011} reported the hard TeV $\gamma$-ray emission inside the source with an integrated energy flux of 1-10 TeV  1.12 $\times$ $10^{-12}$ erg $\rm cm^{-2}$ $\rm s^{-1}$.
\citet{Abdalla2018a} verified that the TeV $\gamma$-ray emission was mainly from the PWN G327.15-1.04 within COMP G327.1-1.1 based on both the coincidence of a spatial position and the approximate size of the multiwavelength emission region. 
In fact, the X-ray and radio observations have also revealed that the morphology of the central PWN is shaped by reverse shock interactions inside the SNR \citep{Temim2009,Temim2015,Ma2016}, as reverse shocks inside SNRs can accelerate electrons to high enough energies to produce gamma-ray emission. 
However,
the pulsed emission of the central  putative pulsar has not been detected from the radio to high energy $\gamma$-ray bands so far.

Compared with other composite SNRs, COMP G327.1-1.1 is an excellent source to explore the interaction between its host and inside PWN in combination with the multiwavelength observations from radio to $\gamma$-ray \citep{Acero2013}. 
In previous works \citep{Acero2013,Acero2016}, the GeV $\gamma$-ray emission from the source had not been detected by the \textit{Fermi}-LAT. In the paper, we search for the potential GeV emission from COMP G327.1-1.1 with the \textit{Fermi}-LAT to better understand its nature of emission in high energy band.

The paper is organized as follows. The routines of data analysis are in Section 2, and the related results are presented in Section 3.  The likely origin of the GeV radiation is discussed in Section 4, and we conclude in Section 5.

\section{Data Reduction} \label{sec:data-reduction}

The analysis was performed using the Femi Science Tools version {\tt v11r5p3}\footnote{http://fermi.gsfc.nasa.gov/ssc/data/analysis/software/}.
We follow the data analysis method as documented in \textit{Fermi} Science Support Center\footnote{http://fermi.gsfc.nasa.gov/ssc/data/analysis/scitools/}.
The Pass 8 data with ``Source'' event class (evtype = 3 \& evclass = 128) and the instrumental response function (IRF) ``P8R3{\_}SOURCE{\_}V2'' were adopted. To minimize the contamination from the Earth Limb, we excluded events with the zenith angle $>$ $90^{\circ}$.  
The time range of the photon events is from August 4, 2008 to October 24, 2020, and the energy range is from 800 MeV to 500 GeV considering a larger point spread function (PSF) in the lower energy band. 
The center of the region of interest (ROI) of $20^{\circ}\times 20^{\circ}$ square is at (R.A. = 238$^{\circ}$.63, Decl. = -55$^{\circ}$.06)\footnote{http://simbad.u-strasbg.fr/simbad/}. 
We used the \textit{Fermi} Large Telescope Fourth Source Catalog \citep[4FGL;][]{4FGL} and the script make4FGLxml.py\footnote{https://fermi.gsfc.nasa.gov/ssc/data/analysis/user/} to generate the source model file, which totally included 737 objects within 30 degrees centered at the center of ROI,
then we added a point source with a power-law spectral model in the position of COMP G327.1+1.1. The fit was performed by using the tool {\tt gtlike} in the \textit{Fermi} Science Tools package\footnote{https://fermi.gsfc.nasa.gov/ssc/data/analysis/scitools/references.html}. Here, we select to free the normalizations and  spectral parameters from all sources within $5^{\circ}$, and the normalizations of the two diffuse backgrounds including {\tt iso{\_}P8R3{\_}SOURCE{\_}V2{\_}v1.txt} and {\tt gll{\_}iem{\_}v07.fits}\footnote{http://fermi.gsfc.nasa.gov/ssc/data/access/lat/BackgroundModels.html}.

\section{Results} \label{sec:data}
\begin{figure}[!h]
  \includegraphics[width=60mm,height=60mm]{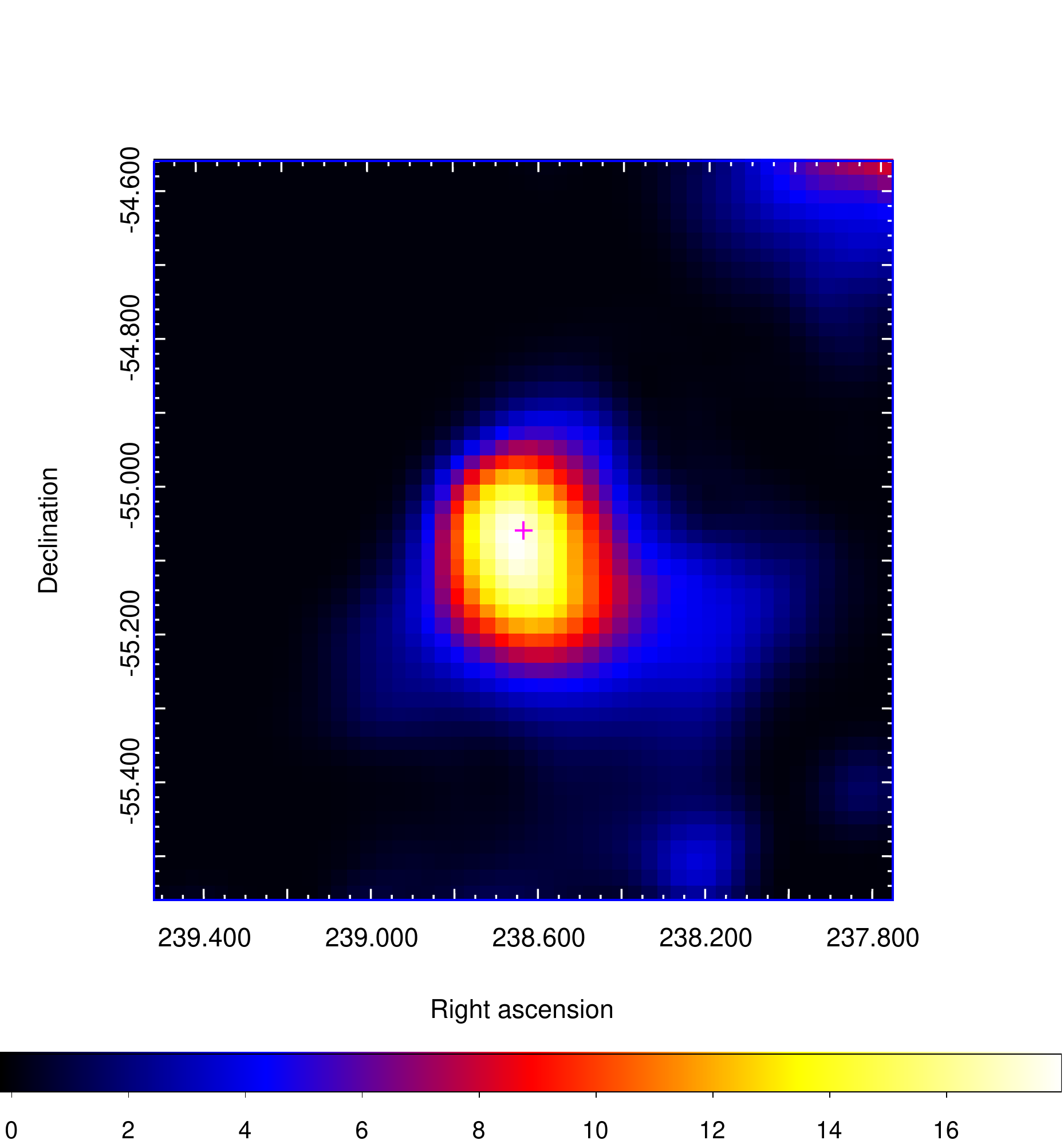}
  \includegraphics[width=60mm,height=60mm]{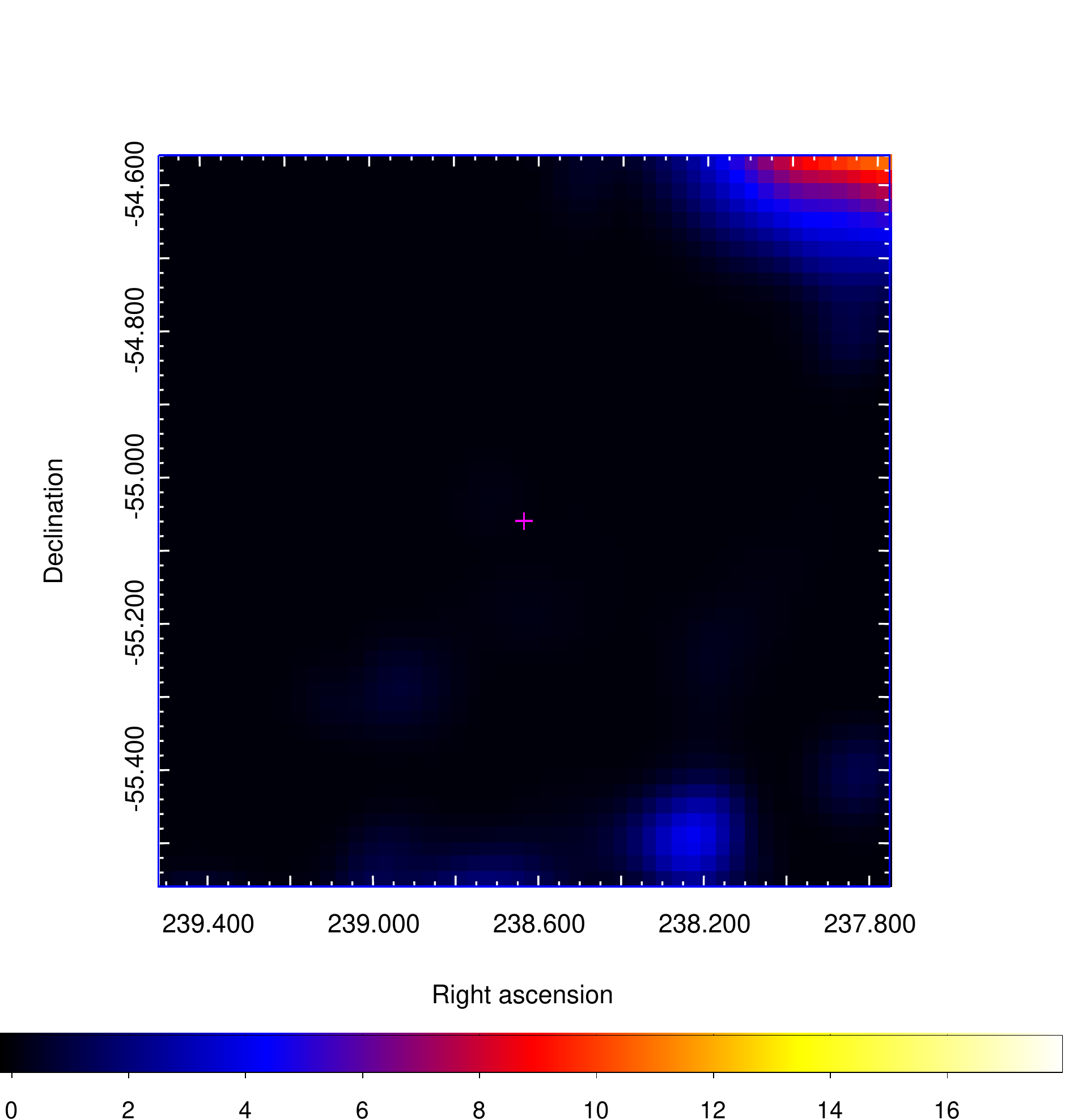}
  \includegraphics[width=60mm,height=60mm]{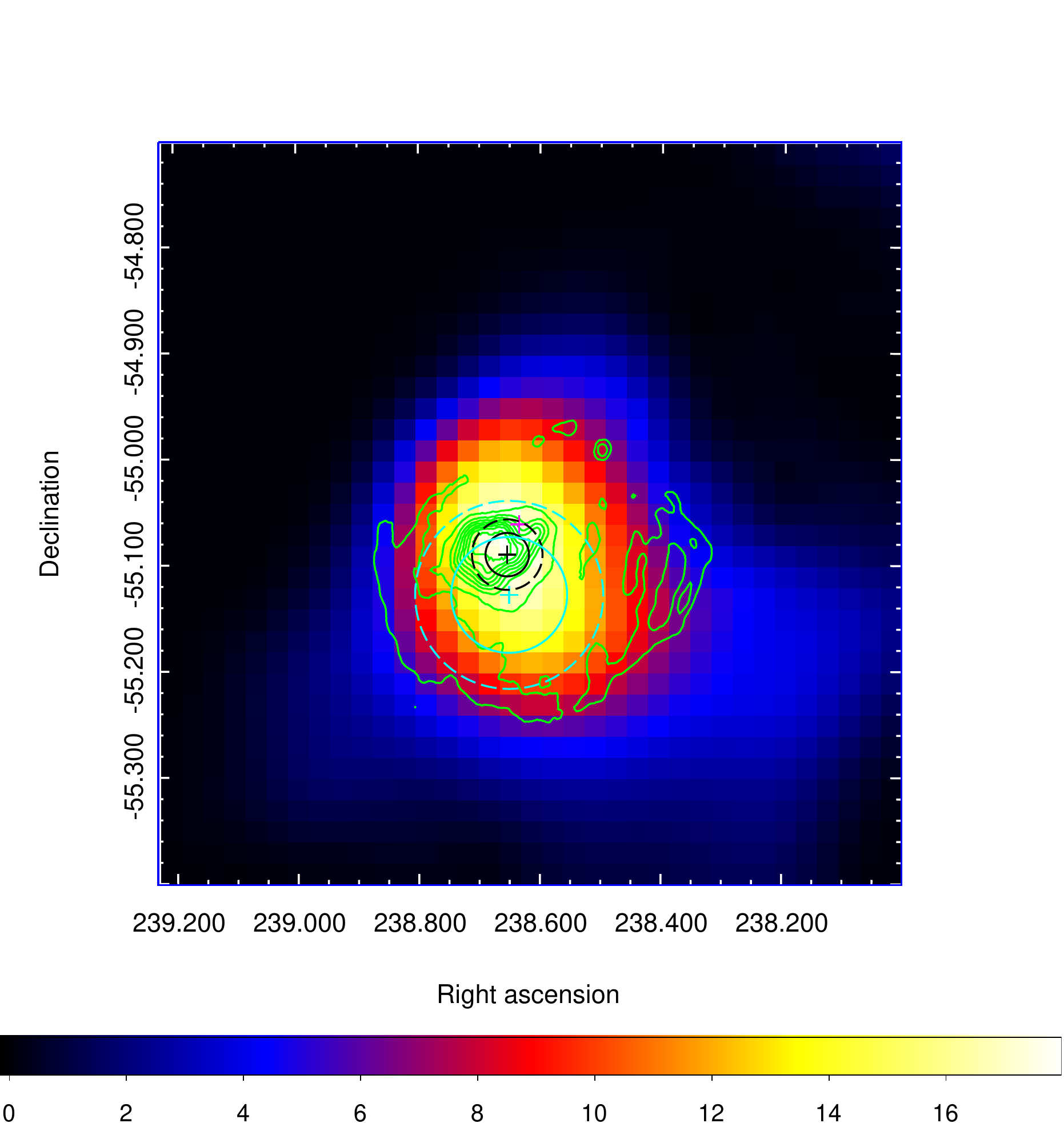}
 \caption{
 TS maps in 0.8-500 GeV with 0$^{\circ}$.02 pixel size centered at the position of COMP G327.1+1.1 marked as a magenta cross. 
Left panel: TS map including all sources from 4FGL in $1^{\circ}.0\times1^{\circ}.0$ region.
Middle panel: TS map subtracted COMP G327.1+1.1 in $1^{\circ}.0\times1^{\circ}.0$ region.  Right panel: TS map in $0^{\circ}.7\times0^{\circ}.7$ region, where the green contours are from observation of MOST at radio 843 MHz, and the cyan circles show the 1$\sigma$ and 2$\sigma$ error regions of the best-fit position marked as a cyan cross from this work. The black circles show the 1$\sigma$ and 2$\sigma$ error regions of the best-fit position marked as a black cross from observation of the H.E.S.S. I
array of Cherenkov telescopes \citep{Abdalla2018a}. A Gaussian function with the kernel radius of $0^{\circ}.3$ was used to smooth these maps.}
    \label{Fig1}
\end{figure}

The commend {\tt gttsmap} is used to compute Test Statistic (TS) map of 1$^{\circ}\times$ 1$^{\circ}$ centered at the position of COMP G327.1+1.1 in 0.8-500 GeV.  We find significant $\gamma$-ray radiation with TS=22.94 from the direction of COMP G327.1+1.1 as shown in the left panel of Figure \ref{Fig1}. Considering the smaller PSF in $\geq$ 0.8 GeV \citep{4FGL}, we do not consider deducting other unknown $\gamma$-ray residual emission more than 1$^{\circ}$.0 in our analysis\footnote{Note: the average containment angle (68\%) over all data of the \textit{Fermi}-LAT PSF is $\sim1^{\circ}$ at 800 MeV \citep{4FGL}.}.
 After subtracting the point source as well, there is no other significant residual $\gamma$-ray radiation from this location as shown in the middle panel of Figure \ref{Fig1}. Therefore, the $\gamma$-ray emission is likely to come from the region.
Next, the best-fit position (R.A., decl.=238.65, -55.13, with 1$\sigma$ error radius of 0$^{\circ}$.04) from COMP G327.1+1.1 was calculated by running {\tt gtfindsrc}, and it is adopted to replace the original position of COMP G327.1+1.1, the new
$\gamma$-ray source is marked as SrcX for all subsequent analyses. 
We found that the location of COMP G327.1+1.1 is within 2$\sigma$ error radius, but the TeV counterpart HESS J1554-550 associated with it is within 1$\sigma$ error radius \citep{Abdalla2018a}, and most of its contours of the brighter center of 843-MHz radio band are within 2$\sigma$ error radius as shown in the right panel of Figure \ref{Fig1}, which implies that the SrcX is likely to be a counterpart of COMP G327.1+1.1.

To check whether SrcX is an extended source, we fit the spatial distribution of the $\gamma$-ray emission from the source using uniform disk models.
The radius for the disk model was set in a range
of 0$^{\circ}$.01-0$^{\circ}$.05 with a step of 0$^{\circ}$.01. 
The fitting results are shown in Table \ref{table1}, and we did not find any significant 
improvement from the uniform disk model by judging TS$_{\rm ext}$ values $\approx$ 0 calculated from 2log($L_{\rm ext}/L_{\rm ps}$) \citep[e.g.,][]{Lande2012}. Here $L_{\rm ext}$ and $L_{\rm ps}$ represent the likelihood values for 
 extended uniform disk with radius of 0$^{\circ}$.01 and point source, respectively. 
  We used the point source template to do all the subsequent analyses.

\begin{table}[!h]
\begin{center}
\caption{Spatial Distribution Analysis for SrcX with two types of spatial models in 0.8-500 GeV}
\begin{tabular}{lcccccc}
  \hline\noalign{\smallskip}
    \hline\noalign{\smallskip}
    Spatial Model & Radius ($\sigma$) & Spectral Index  & Photon Flux & TS
    Value & Degrees of Freedom  \\
                  &     degree     &      &    $\rm 10^{-10}  ph$ $\rm cm^{-2} s^{-1}$  &       & \\
  \hline\noalign{\smallskip}
   Point source    & ...             & 2.35$\pm$0.24 & $5.21\pm1.52$ &22.94   &   4  \\
  \hline\noalign{\smallskip}
   uniform disk    & 0$^{\circ}$.01  & 2.61$\pm$0.07 & $9.19\pm 0.74$ &22.06 &5     \\
       & 0$^{\circ}$.02  & 2.61$\pm$0.07 & $9.18\pm 0.74$ &22.02 &5     \\
       & 0$^{\circ}$.03  & 2.60$\pm$0.07 & $9.17\pm 0.74$ &21.96 &5     \\
       & 0$^{\circ}$.04  & 2.59$\pm$0.07 & $9.16\pm 0.74$ &21.87 &5     \\
       & 0$^{\circ}$.05  & 2.58$\pm$0.07 & $9.14\pm 0.74$ &21.76 &5     \\
  \noalign{\smallskip}\hline
\end{tabular}
\end{center}
    \label{table1}
\end{table}

\subsection{Variability Analysis} \label{sec:data-results}
  
\begin{figure}[!h]
\centering
 \includegraphics[width=\textwidth, angle=0,width=140mm,height=60mm]{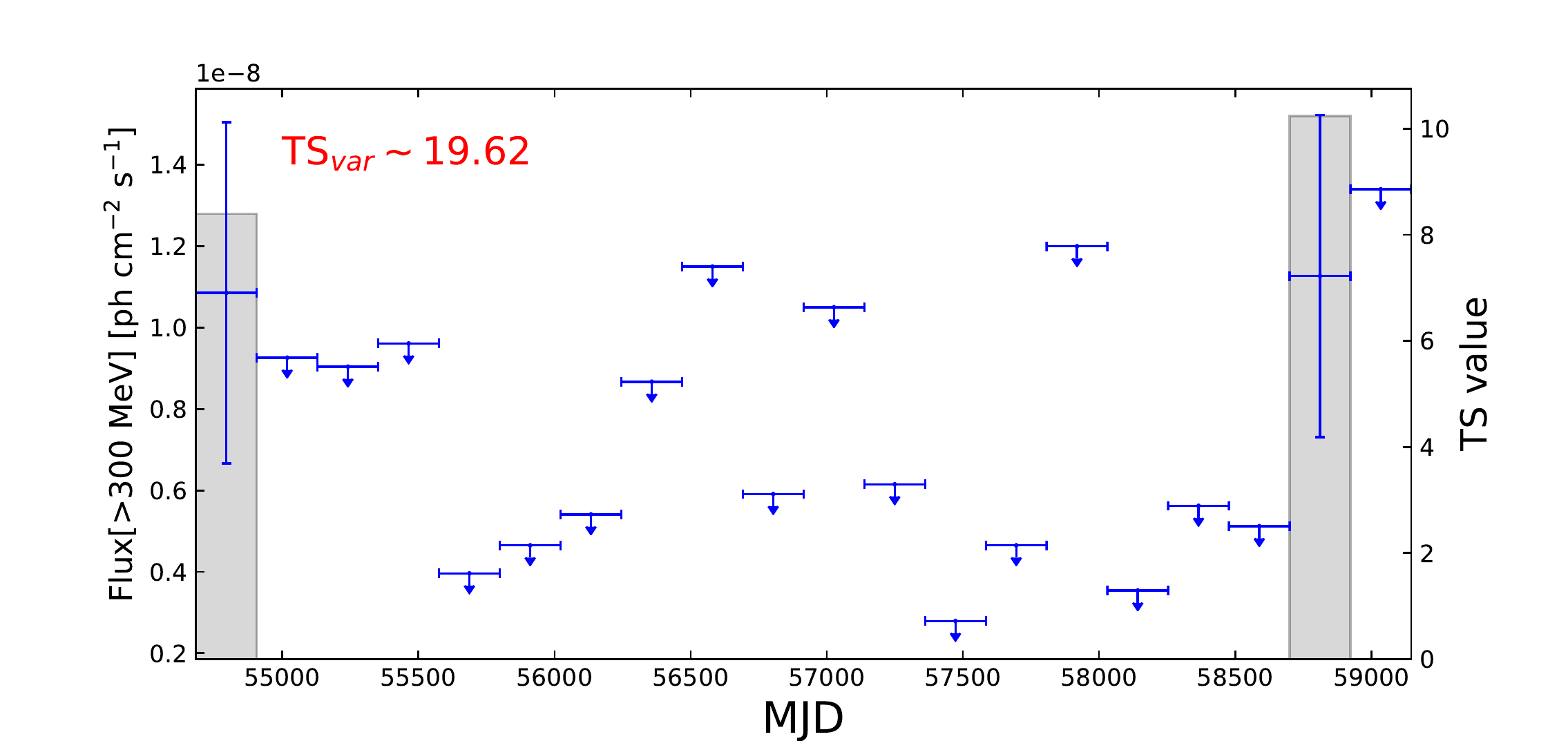} %
 \caption{Light curve of 20-time bins from 0.3-500 GeV for SrcX. 
The upper limits with the 95\% confidence level are given for the time bin of TS value $<$ 4. 
The gray shaded area is used to mark TS value for two bins with TS value $>$ 4.
}
 \label{figure2}
\end{figure}

To check whether SrcX is a variable source, we generate the light curve of $\sim$12.2 years with 20-time bins in 0.3-500 GeV band. Here, we assess the variability by calculating TS$_{\rm var}$ defined by \citet{Nolan2012}.  
For the light curve of 20-time bins, $\rm TS_{var}\geq$ 36.19 is used to identify variable sources at a 99\% confidence level.
However, with a $\rm TS_{var}$=19.62  for SrcX, we do not find any significant variability from Figure \ref{figure2}.

\subsection{\rm Pulsation Search}
The $\gamma$-ray emission signal we detected is only at $\sim$4$\sigma$ confidence level, which is lower than most of the confirmed $\gamma$-ray pulsars listed in 4FGL (only 2 out of 253 $\gamma$-ray pulsars have detection significance lower than 5$\sigma$), which indicates that this source is unlikely to be detected to have $\gamma$-ray pulsations. Nevertheless, we performed timing analysis to the LAT data of COMP G327.1$+$1.1 region to search for possible $\gamma$-ray pulsations. The LAT events within an aperture radius of 2 degrees in 0.8-10 GeV band were included in the analysis since the $\gamma$-ray pulsations from pulsars usually did not be seen above 10 GeV. We weighted these events by their probabilities of originating from the target source using the tool {\tt gtsrcprob} with the fitted source model obtained above and applied the time-differencing blind search technique \citep{Atwood2006} to these weighted events. The search range of the period derivative over period ($\dot{P}/P$) and the frequency ($\nu$) was 0--1.3$\times$10$^{-11}$ s$^{-1}$ (for Crab pulsar) and 0.5--32 Hz, with the steps of 1.545$\times$10$^{-16}$ s$^{-1}$ and 1.90735$\times$10$^{-6}$ Hz, respectively. No significant $\gamma$-ray pulsations from the source region were detected.

\subsection{\rm Spectral Analysis} \label{sec:data-results}
\begin{figure}[!h]
  \centering\underline{}
   \includegraphics[width=90mm]{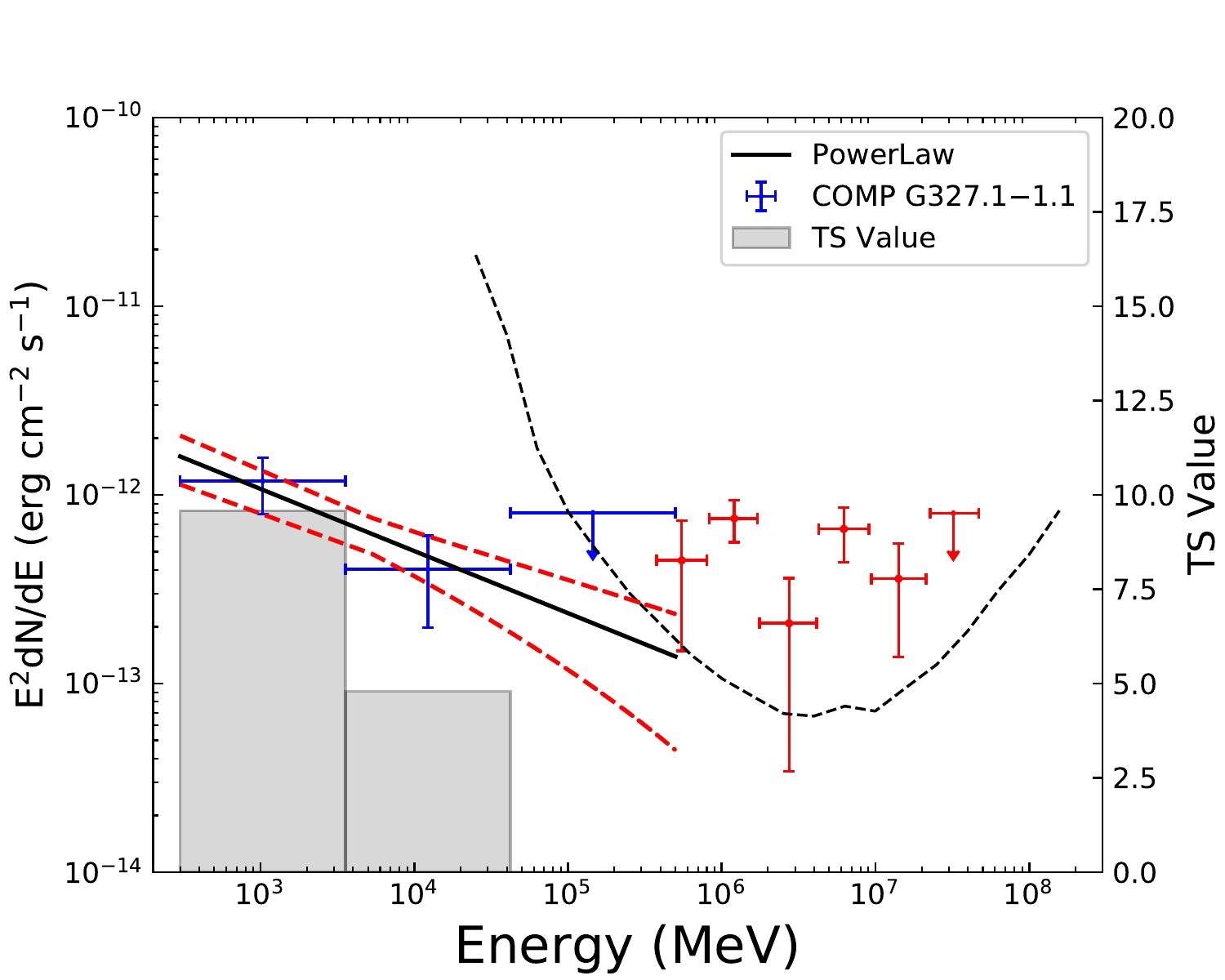}
 \caption{The SED of SrcX in the 0.3-500 GeV band.
 Blue points are \textit{Fermi}-LAT spectral points with the total uncertainty (statistic and systematic).
The black solid and the red dashed lines are used to show the best-fit \textit{Fermi}-LAT spectrum and its 1$\sigma$ statistic uncertainty, respectively. The black dashed line represents the differential sensitivity of CTA-South \citep{CTA}. The red data is from observation of the H.E.S.S. I array of Cherenkov telescopes \citep{Abdalla2018a}. The TS value from each energy bin with the value $>$ 4.0 is represented by using the gray shaded area.  For the third energy bin with TS value $\approx$ 0, the upper limit of 95\% confidence level is given.}
    \label{figure3}
\end{figure}

For the emission of 0.1-0.3 GeV in the position of SrcX, we find there is no significant emission with TS=0 by \textbf{gtlike}. 
The photon flux of the global fit calculated by the binned likelihood analysis method is $\rm (2.51\pm0.31)\times 10^{-9}  ph$ $\rm cm^{-2} s^{-1}$ with the spectral index of $2.44\pm0.07$ in 0.3-500 GeV. We selected the energy band of 0.3-500 GeV to generate its SED.
Here, we choose to divide the energy range of 0.3-500 GeV into three logarithmic bins.
Afterward, each energy bin is separately fitted by using the binned likelihood analysis method, as was done in the global fit.
For the third bin, an upper limit with 95\% confidence level is given for the TS value $\approx$0. The SED of SrcX is shown in Figure \ref{figure3}.  
Meanwhile, we calculate the systematic uncertainty of the first and the second energy bins from the effective area estimated by the bracketing Aeff
method\footnote{https://fermi.gsfc.nasa.gov/ssc/data/analysis/scitools/Aeff{\_}Systematics.html}.  Considering that SrcX is located in the observation range of Cherenkov Telescope Array in the south \citep{CTA}, it will be able to be probed the absence or presence of a cut-off of the $\gamma$-ray spectrum as shown in Figure \ref{figure3}. The data from three bins are shown in Table \ref{table_sed}.

\begin{table}[!h]
\begin{center}
\caption{ The Energy Flux Measurements from SrcX with \textit{Fermi}-LAT}
\begin{tabular}{lccccc}
  \hline\noalign{\smallskip}
    \hline\noalign{\smallskip}
    E  & Band       & $E^{2}dN(E)/dE$ & TS  \\
 (GeV) &  (GeV)     & ($10^{-13}$erg cm$^{2}s^{-1}$ ) &        \\            
  \hline\noalign{\smallskip}
    1.03    & 0.3-35.57    &  11.84$\pm$3.86$^{+0.45}_{-0.71}$ &   9.58  \\
    12.25    & 35.57-42.17    &  4.04$\pm$2.03$^{+0.21}_{-0.37}$  &  4.79      \\
    145.21  & 42.17-500    &  8.04  &  0.0   \\
  \noalign{\smallskip}\hline   
\end{tabular}
    \label{table_sed}
\end{center}
\textbf{Note}: Fluxes with uncertainties from the first and the second energy bins are given with TS$>$4, and the first and the second uncertainties are statistic and systematic ones, respectively. The flux of the third energy bin is the 95\% upper limit.
\end{table}

\section{DISCUSSION} \label{sec:data-results}   

\subsection{\rm Source Detection}
Using the latest \textit{Fermi}-LAT Pass 8 data, we analyzed the $\gamma$-ray radiation from the region of  COMP G327.1-1.1, and found its photon flux was $\rm (9.19\pm2.34)\times 10^{-10}  ph$ $\rm cm^{-2} s^{-1}$  with TS$\approx$22 in 0.8-500 GeV. We analyzed the variability of the light curve of $\sim$12.2 years, and we found that there was no significant variability.
For the higher TS values from the first and the nineteenth bins than others in Figure \ref{figure2}, so we also searched likely weaker active galactic nuclei (AGNs) candidates from the range within the 2$\sigma$ error radius centered at its best-fit position by Aladin\footnote{https://aladin.u-strasbg.fr/aladin.gml} and SIMBAD\footnote{http://simbad.u-strasbg.fr/simbad/}, but we did not find the likely ones. 

 
  We then obtained that its luminosity in 0.3-500 GeV band was \textbf{(9.29$\pm$0.89)$\times 10^{33} \rm erg\ s^{-1}$} with a distance of 4.52 kpc \citep{Wang2020}. We found the magnitude \textbf{range} of the luminosity was consistent with \textbf{those} of the observed galactic COMP SNRs \citep{Liu2015}, and its spectral index of 2.33$\pm$0.12 was also well consistent with the average value \footnote{Note: Here, we exclude the 4 SNRs  (including G0.0+0.0, G132.7+1.3, G156.2+5.7, G290.1-0.8) from Table 3 of \citet{Liu2015} for  the uncertain association with the detected $\gamma$-ray emission.} of 2.34 from 10 COMP SNRs with a power-law spectral model from \citet{Liu2015}.
Moreover, we found that the location of SrcX overlapped well with the location in the 843-MHz radio and 0.2-100 TeV bands, which strongly suggests that SrcX is likely to be a GeV counterpart of COMP G327.1-1.1.

In fact, \citet{Abdalla2018a} had confirmed that the TeV band data of HESS J1554-550 were mainly from the PWN G327.15-1.04 inside COMP G327.1-1.1 according to a spatial coincidence with the PWN identified in radio and X-ray bands and the approximate size of the emission region in radio and X-ray energy bands. Therefore, in combination with the central brighter source structure and the spatial overlap from various energy bands, we suggest that the GeV emission may be powered by the PWN,  and the nature of this emission is likely to be the relativistic electrons from the PWN inverse-Compton upscattering the cosmic microwave background (CMB) and/or SNR photons \citep[e.g.,][]{Zeng2019,Xin2019}. 
By comparing the value of the power-law spectral index$\sim$2.25$\pm$0.20 in 1-100 GeV band of COMP G327.1+1.1 with that of the average power-law spectral index$\sim$2.09 of all PWNs confirmed from 4FGL,  we found that the two are in good agreement, and this result further indicates that its GEV emission probably originates from the internal PWN of COMP G327.1-1.1.
However, the contribution from  shocks of the SNR extended region of COMP G327.1-1.1 for the GeV emission cannot be excluded due to the poor spatial resolution of the \textit{Fermi}-LAT.

\subsection{\rm Model Fitting}
Here, we also combine the following three simple stationary models to explain its SED: 

Model 1: One-zone leptonic model.

Model 2: Two-zone leptonic model.

Model 3: One-zone leptohadronic model.

\citet{Ma2016} suggested that the emission in the KeV and TeV energy bands could respectively originate from synchrotron and inverse-Compton processes for same photon indexes from the two energy bands, so we assume that these models all have the same particle distribution of a power law with an exponential cutoff model (PLEC) for radio and $\gamma$-ray bands.  Here, the form of PLEC is as follows \citep[e.g.,][]{Xing2016,Xin2019}:

\begin{equation}
N(E) = N_{\rm 0}\left (\frac{E}{E_{\rm 0}}\right )^{-\alpha}exp\left (-\frac{E}{E_{\rm cutoff}}\right),
\end{equation}

where N$_{0}$ is the amplitude, $\alpha$ is the spectral index, E is the particle energy, E$_{\rm cutoff}$ is the break energy, and $E_{\rm 0}$ is fixed to 10 TeV.
Here, we use NAIMA \citep{Zabalza2015} to fit multiwavelength spectrum using the Markov Chain Monte Carlo Method in the emcee package \citep{Foreman-Mackey2013}.
The CMB is all considered in the inverse Compton scattering process for the above three models.
For Model 3, the pion production and cross-section of the proton-proton energy losses are selected from PYTHIA 8 \citep{Kafexhiu2014}.

Firstly, we select Model 1 to fit the SED, and we find that this model cannot well explain the high-energy $\gamma$-ray radiation in the GeV band with a reduced $\chi^{2} \sim$2.28.
The discrepancy may imply a more
complicated model for the broadband emission \citep[e.g., multi-emission zones,][]{Xing2016,Lu2020}. 
For the case of multi-emission zones, in fact, \citet{Lu2020} considered a two-zone model with different diffusion processes from the extended region and the PWN to explain the SED of plerionic SNR G21.5-0.9.
Here, for simplicity, we consider Model 2 containing two unknown emission regions to explain the SED and find Model 2 can well explain its GeV high-energy $\gamma$-ray radiation with a reduced $\chi^{2} \sim$1.55. 
However, the hadronic origin cannot be excluded even if the ambient mediums regarded as a natural target from the COMP SNR are rich, and they can provide subsequent $\gamma$-ray production by the interaction of proton-proton collisions \citep[e.g.,][]{Temim2013,Lu2020}. Therefore, we also considered the case of hadrons by using Model 3. Here, we assume the gas density is 10 $\rm cm^{-3}$ \citep[e.g.,][]{Xin2019}.
 With a reduced $\chi^{2} \sim$2.06, we found that the model can also explain the broadband observations well.
The fitting parameters for the above three models are given in Table \ref{table3}.

However, with only fewer data points with a lower confidence level in the GeV band and larger errors in the TeV band, more observations are very necessary to infer the likely origin of the high-energy emission in the future.

\begin{figure}[!h]
  \includegraphics[width=60mm,height=60mm]{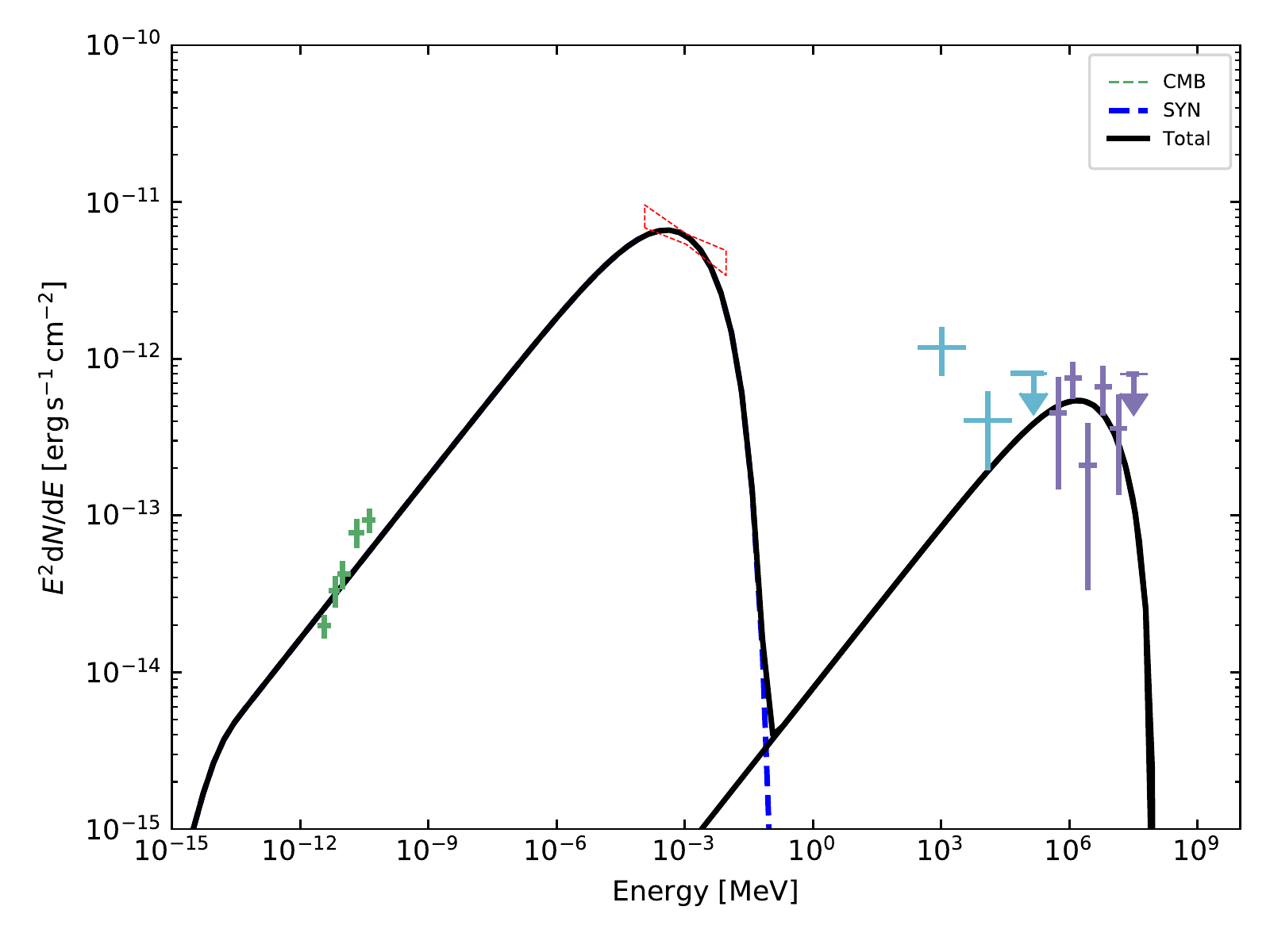}
  \includegraphics[width=60mm,height=60mm]{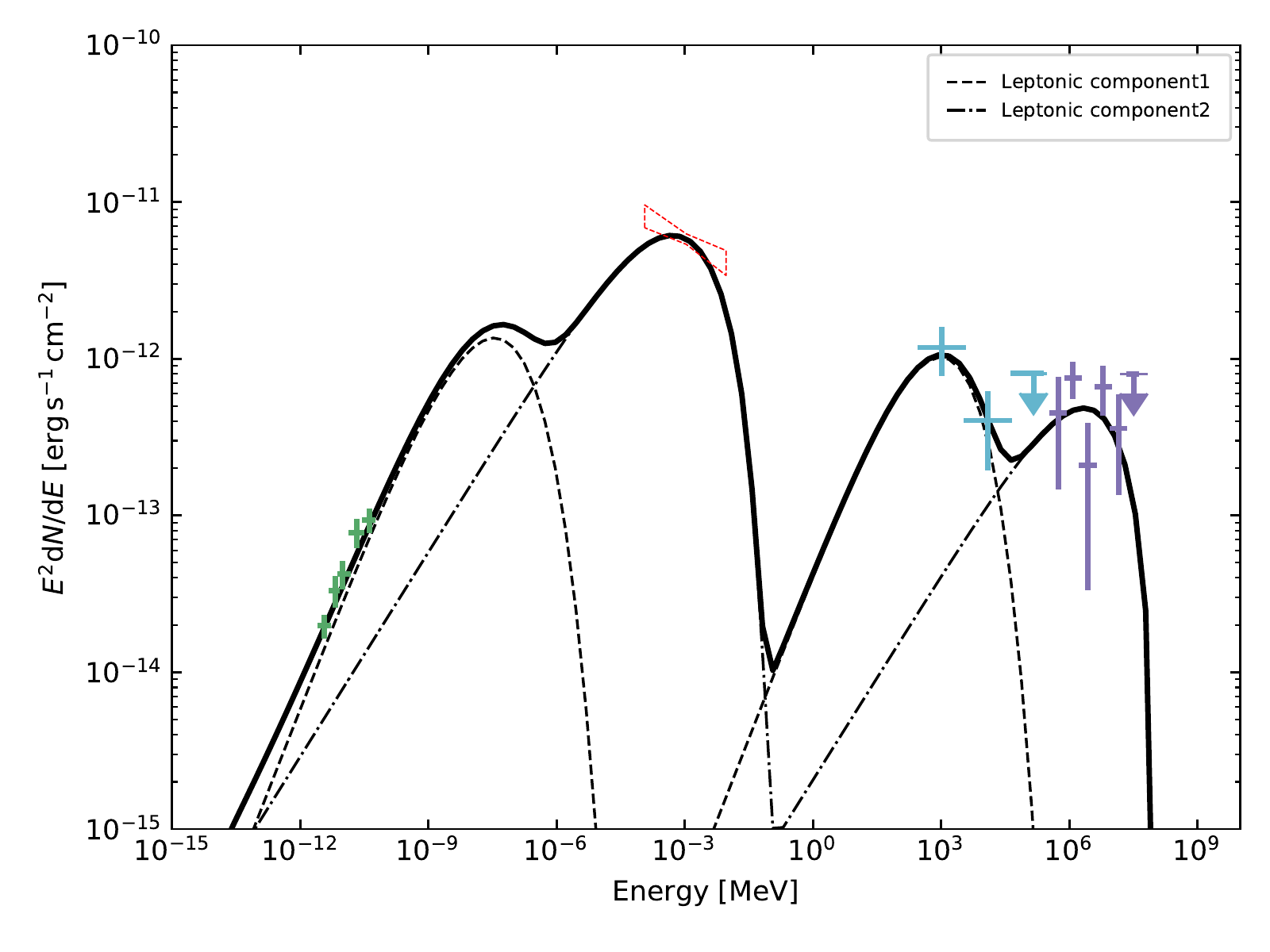}
  \includegraphics[width=60mm,height=60mm]{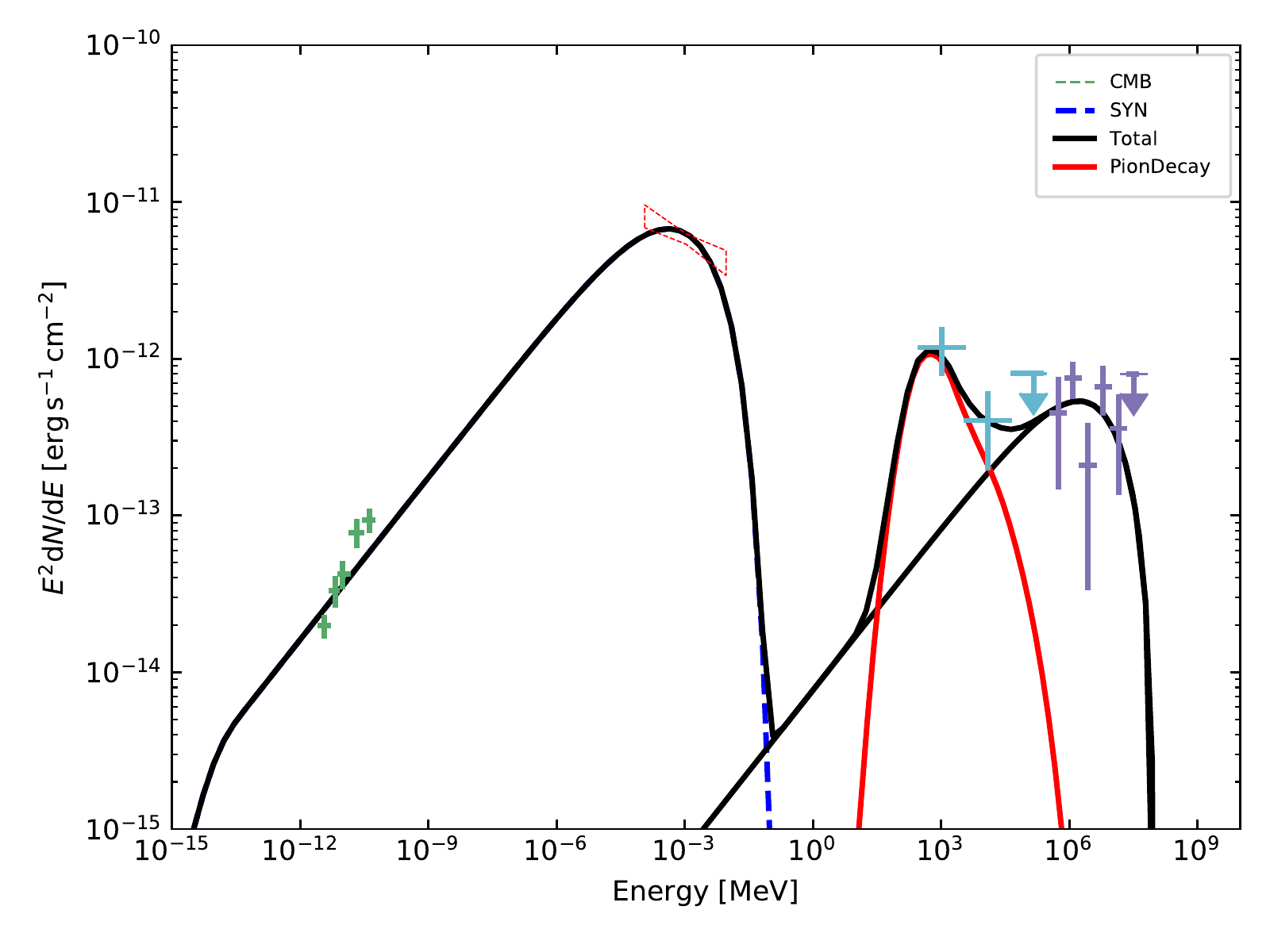}
 \flushleft
 
\caption{
Modeling of multiwavelength spectra of COMP G327.1+1.1. Left panel: the fitting result of Model 1. Middle panel: that of Model 2. Right panel: that of Model 3. Radio data are from \citet{Ma2015}, and X-ray data are from \citet{Temim2015}. For data from GeV to TeV, please see Figure \ref{figure3} for a general description.
}
\label{Fig4}
\end{figure}

\begin{table*}[!h]
\caption{The Best-fit Parameters of Three Models}
\begin{center}
\begin{tabular}{cccccc}
  \hline\noalign{\smallskip}
    \hline\noalign{\smallskip}
  Model Parameter        &  $B$    & $\alpha$               & $E_{\rm cutoff}$       & $W_{\rm e}$ (or $W_{\rm p}$)  & $\chi^{2}/N_{dof}$   \\
                         &  $\mu$G &                        & TeV                    & $10^{48}$ erg   &  \\
  \hline\noalign{\smallskip}
\multicolumn{5}{c}{Model 1} &  $\frac{29.6}{18-5}=2.28$  \\
   \noalign{\smallskip}\hline 
  Leptonic component    & $10.1^{+1.55}_{-1.08}$ & $2.31^{+0.03}_{-0.03}$  & $45.3^{+19.4}_{-11.8}$ & $ 0.67^{+0.17}_{-0.23}$   &     \\
      \noalign{\smallskip}\hline

  \multicolumn{5}{c}{Model 2}  &  $\frac{12.4}{18-10}=1.55$  \\
    \noalign{\smallskip}\hline 
  Leptonic component1    & $4.14^{+1.86}_{-1.28}$ & $1.57^{+0.13}_{-0.11}$  & $0.30^{+0.08}_{-0.09}$ & $ 3.13^{+0.65}_{-0.73}$   &     \\
                     
    \noalign{\smallskip}\hline
  Leptonic component2    & $9.93^{+1.43}_{-1.4}$ & $2.08^{+0.17}_{-0.25}$  & $38.7^{+12}_{-9.09}$ & $ 0.21^{+0.06}_{-0.08}$   &     \\
   \noalign{\smallskip}\hline

     \multicolumn{5}{c}{Model 3} &  $\frac{20.6}{18-10}=2.06$  \\
       \noalign{\smallskip}\hline
  Leptonic component    & $10.2^{+1.62}_{-1.03}$ & $2.31^{+0.03}_{-0.03}$  & $45^{+9.8}_{-9.51}$ & $ 0.73^{+0.05}_{-0.07}$      \\
    \hline\noalign{\smallskip}
  Hadronic component    &  ---                   &$2.65^{+0.10}_{-0.19}$  & $1^{+0.5}_{-0.5}$ & $9.90^{+0.42}_{-0.75}$     \\
  \noalign{\smallskip}\hline  
\end{tabular}
\end{center}
\label{table3}
\end{table*}


\section{Conclusion} \label{sec:conclusion}
In this work, we found significant $\gamma$-ray emission of 0.8-500 GeV from the region of COMP G327.1+1.1 with $\sim4\sigma$ significance level by using \textit{Fermi}-LAT Pass 8 data. Meanwhile, we analyzed the variability from its light curve of $\sim$12 years, and no significant variability is found with $\rm TS_{var}=19.62$.  Its spatial position well coincides with those at the 843-MHz radio and the 0.2-100 TeV energy band.
Moreover, the ranges of the spectral index and the GeV luminosity for the new $\gamma$-ray source are well consistent with the observed COMP SNRs in the Milky Way. We believe that the emission is probably from COMP G327.1+1.1.     
On this basis, in combination with the central brighter structure, we suggest the GeV emission may be powered by the PWN.
 By comparing three radiation models, we found that the two-zone leptonic model can better explain its multiwavelength data.
More data in the high-energy band are necessary to firmly confirm the association between the $\gamma$-ray source and COMP G327.1+1.1 in the future, which will help us better understand its emission origin and acceleration mechanism.

\section{Acknowledgments} 
We sincerely appreciate the referee for his/her invaluable
comments, and we gratefully acknowledge Xian Hou from Yunnan Observatories of Chinese Academy of Sciences for her generous help.
Meanwhile, we also appreciate the support for this work from
National Key R\&D Program of China under grant No. 2018YFA0404204, the National Natural Science Foundation of China (NSFC U1931113, U1738211) ,the Foundations of Yunnan Province (2018IC059, 2018FY001(-003)), the Scientific research fund of Yunnan Education Department (2020Y0039).



\begin{thebibliography}{}


\bibitem[Abdalla et al.(2018)]{Abdalla2018a}Abdalla, H., Abramowski, A., Aharonian, F., et al. 2018a, A\&A, 612, A1

\bibitem[Abdollahi et al.(2020)]{4FGL} Abdollahi, S., Acero, F., \& Ackermann, M. et al. 2020, ApJS, 247, 33

\bibitem[Acero et al.(1991)]{Acero2013}Acero, F., Ackermann, M., Ajello, M., et al. 2013, ApJ, 773, 77

\bibitem[Acero et al.(2016)]{Acero2016}Acero, F., Ackermann, M., Ajello, M., et al. 2016, ApJS, 224, 8

\bibitem[Acero et al.(2011)]{Acero2011}Acero, F., Djannati-Ata\"{\i}, A., F\"{o}rster, A., et al. 2011, ICRC (Beijing), 32, 185

\bibitem[Atwood et al.(2006)]{Atwood2006} Atwood, W.~B., Ziegler, M., Johnson, R.~P., et al.\ 2006, \apjl, 652, L49

\bibitem[Bocchino \& Bandiera (2003)]{Bocchino2003}Bocchino, F., \& Bandiera, R. 2003, A\&A, 398, 195

\bibitem[Clark et al.(1973)]{Clark1973}Clark, D. H., Caswell, J. L., \& Green, A. J. 1973, Natur, 246, 28

\bibitem[Clark et al.(1975)]{Clark1975}Clark, D. H., Caswell, J. L., \& Green, A. J. 1975, AuJPA, 37, 1

\bibitem[CTA Consortium(2019)]{CTA}CTA Consortium, Acharya, B. S., Agudo, I., et al. 2019, Science with the Cherenkov Telescope Array (Singapore: World Scientific)           

\bibitem[Foreman-Mackey et al. (2013)]{Foreman-Mackey2013}Foreman-Mackey, D., Hogg, D. W., Lang, D., et al. 2013, PASP, 125, 306

\bibitem[Kafexhiu et al.(2014)]{Kafexhiu2014}Kafexhiu, E., Aharonian, F., Taylor, A. M., \& Vila, G. S. 2014, Phys. Rev., D90, 123014

\bibitem[Lamb \& Markert(1981)]{Lamb1981}Lamb, R. C., \& Markert, T. H. 1981, ApJ, 244, 94

\bibitem[Lande et al.(2012)]{Lande2012}Lande, J., Ackermann, M., Allafort, A., et al. 2012, ApJ, 756, 5

\bibitem[Lang et al.(2016)]{Lang2013}Lang, D., et al. 2013, PASP, 125, 306

\bibitem[Liu et al.(2015)]{Liu2015}Liu, B., Chen, Y., Zhang, X., et al. 2015, ApJ, 809, 102


\bibitem[Lu et al.(2020)]{Lu2020}Lu, F.-W., Gao, Q.-G., Zhang,L. 2020, ApJ, 889, 30

\bibitem[Ma et al. (2015)]{Ma2015}Ma, Y. K., Ng, C.-Y., Bucciantini, N., et al. 2015, in American Astronomical Society Meeting Abstracts, 225, \#445.06

\bibitem[Ma et al.(2016)]{Ma2016}Ma, Y. K., Ng, C.-Y., Bucciantini, N., et al. 2016, ApJ, 820, 100

\bibitem[Nolan et al.(2012)]{Nolan2012}Nolan, P. L., Abdo, A. A., Ackermann, M., et al. 2012, ApJS, 199, 31


\bibitem[Seward et al.(1996)]{Seward1996}Seward, F. D., Kearns, K. E., \& Rhode, K. L. 1996, ApJ, 471, 887

\bibitem[Sun \& Chen (1999)]{Sun1999}Sun, M., Wang, Z.-r., \& Chen, Y. 1999, ApJ, 511, 274

\bibitem[Temim et al.(2009)]{Temim2009}Temim, T., Slane, P., Gaensler, B. M., Hughes, J. P., \& van der Swaluw, E. 2009, ApJ, 691, 895

\bibitem[Temim et al. (2013)]{Temim2013}Temim, T., Slane, P., Castro, D., et al. 2013, ApJ, 768, 61

\bibitem[Temim et al.(2015)]{Temim2015}Temim, T., Slane, P., Kolb, C., et al. 2015, ApJ, 808, 100

\bibitem[Wang et al.(2020)]{Wang2020}Wang, S., Zhang, C.-Y., Jiang, b.-w., et al.  2020, A\&A, 639, A72

\bibitem[Whiteoak \& Green(1996)]{Whiteoak1996}Whiteoak, J. B. Z., \& Green, A. J. 1996, A\&AS, 118, 329

\bibitem[Xin et al. (2019)]{Xin2019}Xin, Y.-L., Zeng, H.-D., Liu, S.-M., Fan, Y.-Z., \& Wei, D.-M. 2019, ApJ, 885, 162

\bibitem[Xing et al.(2016)]{Xing2016}Xing, Y., Wang, Z.-X., Zhang, X., Chen, Y. 2016, ApJ, 823, 44

\bibitem[Zabalza (2015)]{Zabalza2015}Zabalza, V. 2015, in press (arXiv:1509.03319)

\bibitem[Zeng (2019)]{Zeng2019}Zeng, H., Xin, Y., \& Liu, S. 2019, ApJ, 874, 50

\end{thebibliography}
\end{document}